\newcommand{\ms}{\si{\milli\second}}%
\newcommand{\PaperTitle}{
Ultrafast Image Categorization in Biology and Neural Models
}
\newcommand{\AuthorJN}{Jean-Nicolas J\'{e}r\'{e}mie}
\newcommand{\AuthorLP}{Laurent U Perrinet}

\newcommand{\Abstract}{
Humans are able to categorize images very efficiently, in particular to detect the presence of an animal very quickly. Recently, deep learning algorithms based on convolutional neural networks (CNNs) have achieved higher than human accuracy for a wide range of visual categorization tasks. However, the tasks on which these artificial networks are typically trained and evaluated tend to be highly specialized and do not generalize well, e.g., accuracy drops after image rotation. In this respect, biological visual systems are more flexible and efficient than artificial systems for more general tasks, such as recognizing an animal. To further the comparison between biological and artificial neural networks, we re-trained the standard VGG 16 CNN on two independent tasks that are ecologically relevant to humans: detecting the presence of an animal or an artifact. We show that re-training the network achieves a human-like level of performance, comparable to that reported in psychophysical tasks. In addition, we show that the categorization is better when the outputs of the models are combined. Indeed, animals (e.g., lions) tend to be less present in photographs that contain artifacts (e.g., buildings). Furthermore, these re-trained models were able to reproduce some unexpected behavioral observations from human psychophysics, such as robustness to rotation (e.g., an upside-down or tilted image) or to a grayscale transformation. Finally, we quantified the number of CNN layers required to achieve such performance and showed that good accuracy for ultrafast image categorization can be achieved with only a few layers, challenging the belief that image recognition requires deep sequential analysis of visual objects. We hope to extend this framework to biomimetic deep neural architectures designed for ecological tasks, but also to guide future model-based psychophysical experiments that would deepen our understanding of biological vision.
}
\newcommand{\Keywords}{vision; ultrafast animal categorization; deep learning; transfer learning; computational neuroscience; behavior; image categorization; timing}
\newcommand{\Acknowledgments}{For the purpose of open access, the author has applied a CC BY public copyright licence to any author accepted manuscript version arising from this submission.} %
\newcommand{\DataAvailability}{
This work is made reproducible using the following tools. First, the code reproducing all figures is {available at} 
 \href{https://github.com/SpikeAI/2022-09_UltraFastCat/blob/main/Readme.md}{GitHub}~\citep{Jeremie2022a} (accessed on 15 March 2023), and in particular the {code at} 
 \href{https://github.com/SpikeAI/DataSetMaker}{DataSetMaker}~\citep{Jeremie2022b} (accessed on 15 March 2023) was used to retrieve images. The paper is available {as an} 
 \href{https://arxiv.org/abs/2205.03635}{arXiv preprint} with links to previous versions and to the code (accessed on 15 March 2023). Also find the {associated} 
 \href{https://www.zotero.org/groups/4560566/ultrafastcat}{zotero group} (accessed on 15 March 2023) used to regroup relevant literature on the subject.

}

\documentclass[vision,article,accept, 
pdftex,moreauthors]{Definitions/mdpi} 
\firstpage{1} 
\pubvolume{1}
\issuenum{1}
\articlenumber{0}
\pubyear{2023}
\copyrightyear{2023}
\datereceived{30 September 2022} 
\daterevised{9 March 2023} 
\dateaccepted{15 March 2023} 
\datepublished{} 
\hreflink{https://doi.org/} 

\usepackage[position=top,labelformat=simple]{subfig} 
\usepackage{siunitx}
\usepackage{amsmath,amssymb,amsfonts}
\usepackage{graphicx}

\usepackage[disable]	
	{todonotes}

\newcommand{\noteLP}[1]{\todo[color=blue!20, inline, author=\textbf{LP}]{#1}}

\Title{\PaperTitle}

\TitleCitation{\PaperTitle}

\Author{\AuthorJN~{*}
\orcidA{}, {\AuthorLP} 
~{*}\orcidB{}}

\AuthorNames{\AuthorJN and \AuthorLP}

\AuthorCitation{J\'{e}r\'{e}mie, J.-N.; Perrinet, L.U.}

\address[1]{%
Institut de Neurosciences de la Timone (UMR 7289), Aix Marseille {Univ,} 
 {CNRS,} 
 {13005 Marseille}
, France
}

\corres{Correspondence: jean-nicolas.jeremie@univ-amu.fr (J.-N.J.); laurent.perrinet@univ-amu.fr (L.U.P.)}

\abstract{\Abstract}


\begin{document}


\section{Introduction}
\unskip
\subsection{Biological Vision and Ultrafast Image~Categorization}
What distinguishes a visual scene that includes an animal from one that does not? This question of ``animacy detection'' is crucial for the survival of any species, especially in regard to the interactions between prey and predators. This constraint has therefore profoundly shaped the way biological visual systems process retinal input. Of~particular importance is the fact that this response must be efficient and fast, while keeping energy requirements to a minimum. In~addition, these systems must fit the ecological niche of the system under consideration, with~the range of patterns to be recognized being different for, say, a~lion, a~bird, and~a human. It is important to note that this biologically-inspired approach shares some similarities and differences with detection algorithms defined in computer vision. Our goal in this paper is both to propose bio-inspired ultrafast image categorization models and to better understand how biological visual systems can efficiently implement such a task~\citep{Cristobal2015}.

\begin{figure}[H]
\subfloat[\centering{Synset Animal \\ - Target -}]{\includegraphics[width=.25\linewidth]{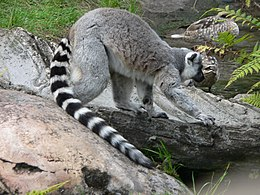}}\hspace{10pt}%
\subfloat[\centering{Synset Animal \\ - Distractor -}]{\includegraphics[width=.25\linewidth]{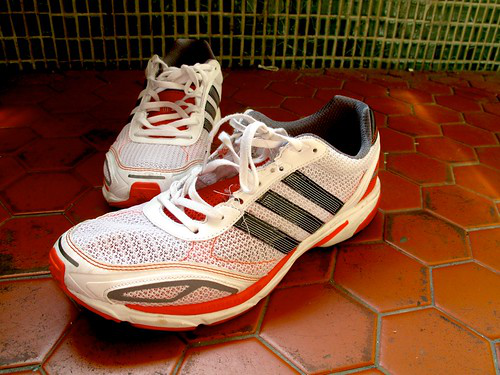}}%
\\
\subfloat[\centering{Synset Artifact \\ - Target -}]{\includegraphics[width=.25\linewidth]{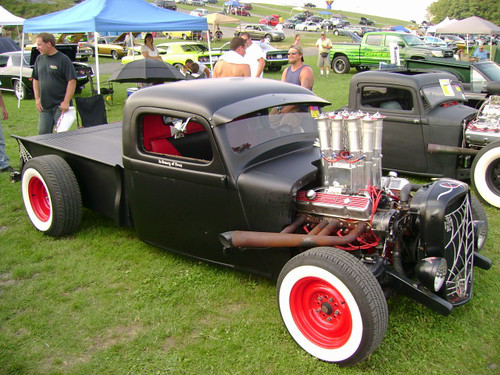}}\hspace{10pt}%
\subfloat[\centering{Synset Artifact \\ - Distractor -}]{\includegraphics[width=.25\linewidth]{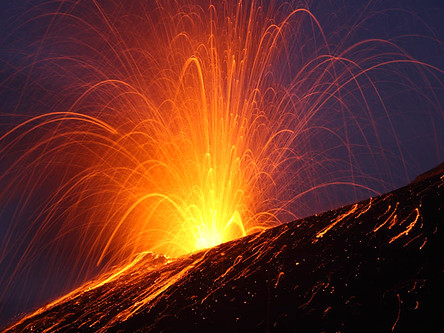}}
\caption{
In ultrafast image categorization, the~task is to report whether a briefly flashed image contains a class of object, such as an animal~\citep{Thorpe1996}. The~presentation time can be on the order of $20~\ms$, and the response is, for~example, the~pressing or not pressing of a button. Representative images for distractors and targets are shown here for two classes: `animal' and `artifact'. Note that these tasks are {a priori} independent and that an animal target can be either a target or a distractor for the other task. Here, based on~\citet{Rousselet2003}, we did not consider images of humans to be part of the animal class, since they seem to represent a class of their own.
}
\label{fig:figure_1}
\end{figure}

Therefore, let us first define the task of rapidly detecting an animal in a scene (see Figure~\ref{fig:figure_1}). This task is routinely used in the study of biological vision in the laboratory (for a review, see~\citep{Fabre-Thorpe2011}). In~its simplest form, it consists of reporting whether an image contains an animal. When applied to generic natural scenes, the~task is such that the animal species is arbitrary and can include, for~example, birds, insects, or~mammals. A~further difficulty is that there are large variations in the identity, shape, pose, size, number, and~position of animals that may be present in the scene. However, biological visual systems are able to perform such detection efficiently in briefly flashed images, a~so-called tachistoscopic presentation, with~differential activity in electroencephalogram (EEG) recordings as low as $120~\ms$ for humans~\citep{Thorpe1996} or as low as $80~\ms$ for monkeys~\citep{Fabre-Thorpe1998}. Such EEG recordings from an ultrafast image categorization task are openly available~\citep{Delorme2021}. This has also been observed in the differential activity of single neurons in the primate lateral prefrontal cortex~\citep{Freedman2001}.

Human categorization of an animal can be performed with high accuracy (generally over $80\%$ correct), very quickly~\citep{Kirchner2006}, and~is robust to geometric transformations~\citep{Rousselet2003}. Color seems to have little effect, but~some low-level statistics~\citep{Mirzaei2013}, as~well as other factors (such as the animal's position and size in the scene) may influence accuracy but not speed~\citep{Zhu2013}. Accuracy is maximal when the animal is in the center of the visual field~\citep{Thorpe2001a}, but~performance is still above chance level (at about $60\%$) at extreme eccentricities of about $70^\circ$. Such a task is performed seamlessly in parallel, so that multiple images can be categorized at once~\citep{Drewes2011}. Surprisingly, once the task is learned, novel images are processed as quickly as familiar ones~\citep{Fabre-Thorpe2001}. Given the difficulty of modeling this task, a~scientific question is to understand what features in the image configuration are sufficient to produce such an effective behavioral response~\citep{Crouzet2011}.

\subsection{Feed-Forward Models of Ultrafast Image~Categorization}
Designing the best algorithm to solve ultrafast image categorization, as~implemented in biological systems, is one possible way to answer this question. In~this case, there are major constraints in the dynamics of vision, especially related to the limits of axonal transduction speed, which can lead to major difficulties in modeling the system~\citep{Perrinet2014}. In~the case of the ultrafast go/no-go categorization task, two consequences follow from these physiological constraints: first, the~response must be made quickly and therefore must be open-loop, i.e.,~before the action can take effect; second, since the whole process involves several processing steps before recurrent loops can refine neural activity, the~flow of information is predominantly feed-forward~\citep{Thorpe2001}. This has been confirmed by EEG recordings of humans performing the task, showing that top-down signals (such as context or expectation) can influence categorization, but~that the process is mainly a bottom-up, feed-forward process~\citep{Delorme2004}. We can also expect that there should be a trade-off between accuracy and speed for image categorization algorithms~\citep{Delorme2010}. 

Given the problem of designing the best algorithm to solve ultrafast image categorization in biologically inspired systems, it was previously shown that such a feed-forward architecture may be sufficient to perform the task~\citep{Serre2007}. This architecture consists of a sequence of layers that interleave a linear and a nonlinear process. This is similar to the simple and complex sublayers observed in the primary visual cortex. The~linear part of the processing is performed by a convolutional operator, hence the name convolutional neural network (CNN) for this class of architecture. The~nonlinear operation is often a simple rectifying unit, similar to the integration process that transforms the analog input to a neuron into a (positively defined) firing rate. In~these architectures, the~layer's resolution generally becomes progressively coarser along the levels of the hierarchy, until~a few classification layers provide the final output~\citep{Lecun1998}. The~efficiency of this model yielded results comparable to humans performing the task on the same images~\citep{Serre2007}. Other popular methods use oriented luminance gradient histograms~\citep{Rangdal2014}, but~with a similar architecture, in which a sequence of processing steps in image space is followed by a classification step. Remarkably, these CNN architectures mirror that of the primate visual system, wherein the retinal image is transmitted from the thalamus to the primary visual cortex and then follows a path along the temporal lobe~\citep{Thorpe2001,Grimaldi2022a}.

\subsection{Related~Work}
Since their adoption as modeling tools, feed-forward architectures have been instrumental in the breakthrough of deep learning architectures, in~particular in providing human-like performance for the PASCAL~\citep{Everingham2015} and {\sc ImageNet}~\citep{Russakovsky2015} challenges, that is, classifying millions of images into over $1000$ different categories (labels). An~important aspect of these architectures, originally inspired by neuroscience, is that they can be trained in a supervised manner, i.e.,~by associating each image with a given label in the training phase. This was illustrated for the MNIST challenge of classifying handwritten digits by associating each image of a digit with its recognized value~\citep{Lecun1998}. This training process optimizes a given loss function applied to each pair, which allows the weights of the network to be progressively adjusted using gradient descent. In~particular, a~CNN such as {\sc Vgg 16} is a well-optimized architecture for performing this challenge of computer domain categorization {\sc ImageNet}~\citep{Simonyan2015}. Therefore, we decided to use {\sc Vgg 16} with  {\sc ImageNet} as a starting point to better understand the process underlying ultrafast categorization of natural images, while bridging our knowledge between neuroscience and computer~science.

The task defined for the {\sc ImageNet}~\citep{Russakovsky2015} challenge could be considered computer-specific, since it requires choosing among $1000$ labels, which implies knowing and remembering these $1000$ labels to make the choice. Unlike artificial neural networks, which can easily compare these $1000$ possibilities simultaneously, one can instead use a subset of behaviorally relevant labels to make the task more relevant to humans. Since we defined a novel task, it is then possible to ``re-train'' these CNNs to categorize images by defining a novel set of supervised pairs (e.g., an~image containing an umbrella associated with the synset ``artifact''). For~the original {\sc ImageNet}~\citep{Russakovsky2015} challenge, each input--output data pair consists of an input data point (an image from the {\sc ImageNet} database) and its corresponding output label (e.g., an~image containing an umbrella associated with the label ``umbrella''). The~idea is to take the knowledge gained from one task and transfer it to a different but related task by using the right training pairs to re-train the CNN; this method is called transfer learning~\citep{Yosinski2014}. The~advantage of using this method is that one can more easily explore the space of all possible architectures by adjusting the synaptic weights of the convolutional kernels, but~also by testing the meta-parameters of the CNNs, such as the number of layers, the~number of channels in each layer, or~the coarsening of the visual information along the hierarchy~\citep{Cichy2016}. Note that, at~the extreme, even the best CNN network may not be able to learn to categorize an image-independent feature, such as whether the calendar day on which the photo was taken is odd or even. For~instance, we will show below that, following that logic, if~we define a task consisting of random labels among the $1000$ categories of {\sc ImageNet}, then none of our tested architectures can learn this task efficiently. Finally, while a drawback of these networks is their lack of interpretability, we will exploit the fact that their raw efficiency gives a lower bound for the possibility of solving a given~task.

Indeed, compared to random labels, the~situation is different when defining more ecological tasks, such as categorizing animals or artifacts. This method can also be used by changing the definition of the supervision pairs to study changes in task context, rapid categorization, and object interference in the image~\citep{Joubert2007}. A~fitting question might be, ``Is there an animal in this image?'', since it reduces this human--machine bias by reducing the choices while maintaining a sufficiently complex and documented question. Searching for these kinds of categories seems to be a primordial function of the brain~\citep{Thorpe2001,Kriegeskorte2008}. For~example, using a set of specific stimuli, it has been shown that categories can be found in the brain areas of rhesus monkeys and that these categories can then be learned by artificial neural networks~\citep{Bao2020}. Our goal here is to obtain a model that is more faithful to the physiological data. In~summary, and~somewhat counterintuitively, compared to biological systems, it may be more difficult for a neural model to make a choice between only two alternatives, such as detecting an animal in an image, than~to choose from $1000$ labels~\citep{Mace2009,Mack2015}. This work will allow us to better understand how this is achieved in both biology and computational neuroscience models. 
\subsection{Main~Contributions}%
\begin{figure}[H]
\includegraphics[width=.6\linewidth]{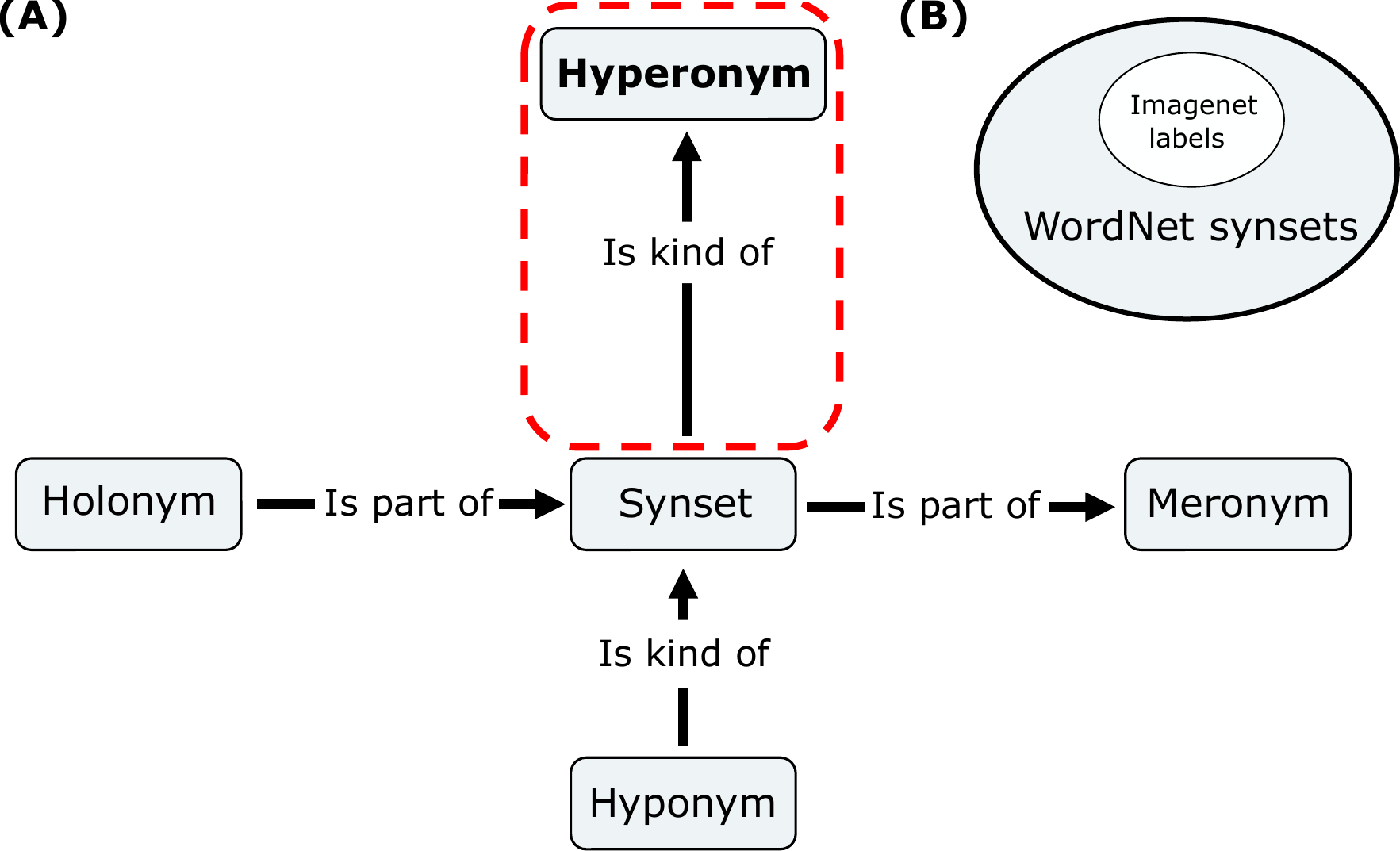}
\caption{
{(\textbf{A})} 
 Schematic displaying the different semantic links between the synsets of the {\sc WordNet} network. We focused on the hyperonym link (frame in red) to build datasets from {\sc ImageNet} synsets; for~instance, which images are `kind of' an `animal'. {(\textbf{B})} Venn diagram exposing the subset of the {\sc ImageNet} labels included in the {\sc WordNet} synset's set.
}
\label{fig:wordnet}
\end{figure}

What distinguishes an image with an animal from an image without an animal? To answer this scientific question, our work proposes three major contributions. First, to~define the psychophysical task, we built a script to build large, arbitrary datasets of images based on {\sc ImageNet}~\citep{Russakovsky2015}. It was defined by selecting labels according to a large semantic graph of English words: {\sc WordNet}~\citep{Fellbaum1998} (see Figure~\ref{fig:wordnet}). According to our scientific question, we first defined our task as the categorization of an animal in an image. As~a control, we also defined an independent task consisting of detecting the presence of any artifact in the image (see Figure~\ref{fig:figure_1}c,d). Second, we re-trained the existing {\sc Vgg 16} model on these tasks and compared its performance with experimental data. This allowed us to test the robustness of our networks to different geometric transformations and to compare their accuracy with that observed in the physiological data. In~addition, we compared the accuracy for both tasks, individually and jointly. Third, we tested different levels of complexity of such models by performing a gradual removal of layers from the original network. This experiment quantified whether low-level features could be sufficient to categorize animals~\citep{Perrinet2015} (although it is known that the global image statistics~\citep{Drewes2005} or the spatial frequency envelope is not sufficient to categorize images~\citep{Wichmann2010}) and whether this could be accompanied by a decrease in invariance to geometric deformations. Finally, we discuss how this work can be useful in the design of future physiological experiments and in the design of novel computer vision architectures.

\unskip

\section{Methods}
\unskip

\subsection{Building the Dataset Maker Library Using the {\sc WordNet} Hierarchy}\label{dataset_maker}
To re-train a deep convolutional network (like {\sc Vgg 16}) for a specific task, one of the most important components is the dataset. We needed a tool that would allow us to generate datasets suitable for answering our question. Therefore, we created a library that, from~a keyword, generates a dataset with image folders containing (target) or not containing (distractor) this keyword~\citep{Jeremie2022b}. For~this, we will use the corresponding set of labels from the {\sc ImageNet} database~\citep{Russakovsky2015}, which is based on a large lexical database of the English language: {\sc WordNet}~\citep{Fellbaum1998}. The~nouns, verbs, adjectives, and~adverbs in this database are grouped into a graphical set of cognitive synonyms, synset, each of which expresses a different concept. These synsets are linked to each other using a few conceptual relations (see Figure~\ref{fig:wordnet}). For~example, if~we set the dataset maker with the keyword `animal', we used the hyperonym link to determine that a German Shepherd is a type of dog and that a dog is a type of `animal', thus defining a hyperonym path. In~this example, the~synset `animal' from  {\sc WordNet} is in the hyperonym path of the label `German Shepherd' in {\sc ImageNet}. Based on this relationship, the~dataset creator selected a specific subset of labels in the {\sc ImageNet} database to build our datasets. Once the list of labels corresponding to our task was selected, the~dataset maker randomly selected from the URLs provided for the {\sc ImageNet} challenge~\citep{Russakovsky2015} to download the images that make up the~dataset. 

With this tool, we generated datasets according to two given tasks. In~particular, we generated the dataset necessary to train the network to answer our question, ``Is there an animal in this scene?'', by selecting the `animal' synset. To~answer the question, ``Is there an artifact in this scene?'', we followed the same protocol. As~a control, we also created a `random' dataset, which was generated by randomly selecting $500$ labels from the {\sc ImageNet} database. The~latter was generated to infer the role of the possible links between arbitrary labels by measuring the resulting efficiency of categorization by a deep convolutional network. In~summary, we used Dataset Maker to generate three datasets: One based on the `animal' synset, one based on the `artifact' synset, and a `random' one. Each newly generated dataset contains a `test', `validation', and~`train' set (with $1200$, $800$, and~$2000$ images, respectively). Each set contained a `target' and a `distractor' category (both with the same number of images). All networks were trained on the `training' set and tested during training on the corresponding `validation' set. We then computed accuracies using the `test' set. As~a control, we also tested the networks on the dataset from~\citet{Serre2007}, which contains a total of $600$ targets (images with an animal) and $600$ distractors (images without an animal). 
%

%
\subsection{Transfer~Learning}
\begin{figure}[H]
\includegraphics[width=.8\linewidth]{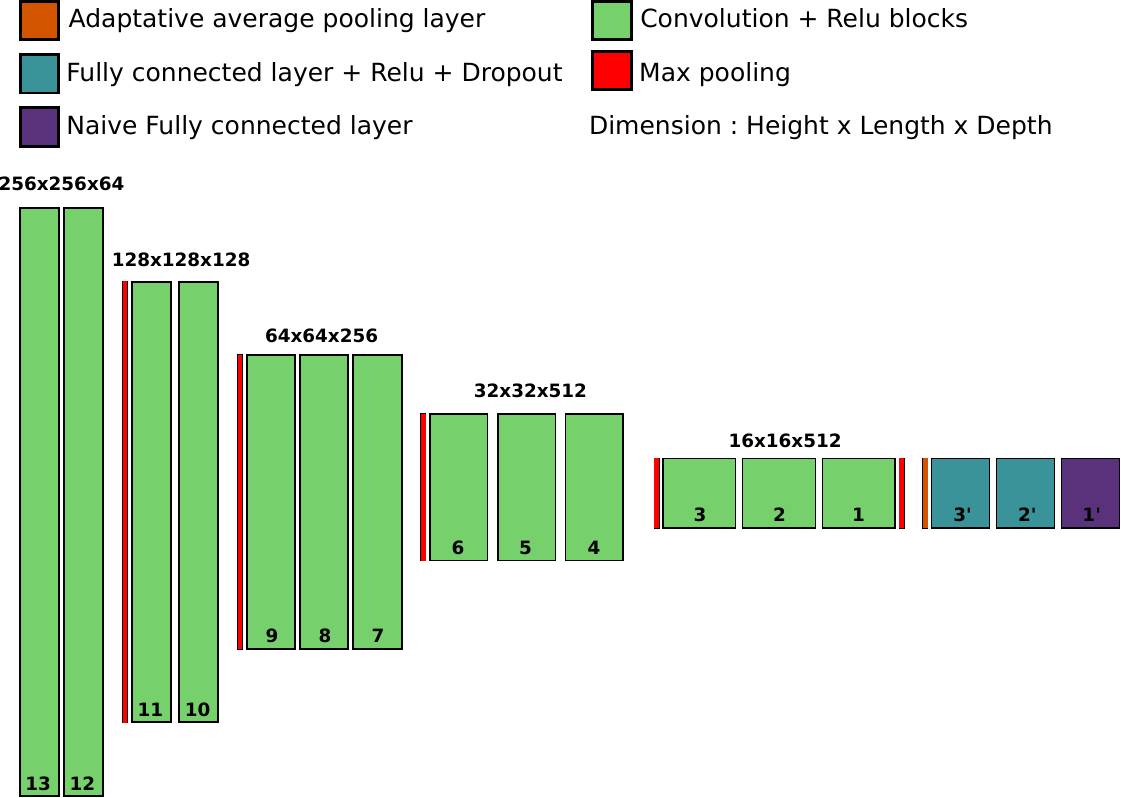}
\caption{
{A diagram} of the network architecture used in transfer learning.  In~green, the~5 blocks of 13 convolutional layers noted from 13, at~the entrance, to~1, at~the junction with the fully connected part, represented jere in blue and purple. The~purple layer represents the naive layer substituting for the one in the {\sc PyTorch} architecture.  }
\label{fig:transfer_learning} 
\end{figure}

We used the transfer learning method to re-train networks~\citep{Yosinski2014}. This method takes the knowledge gained from one task and applies it to a different but related task. We used an existing network that had been pre-trained on a specific task: {\sc Vgg 16}~\citep{Simonyan2015}. This architecture is loaded thanks to the {\sc PyTorch} library~\citep{Paszke2019} and trained on the database used to solve the {\sc ImageNet}~\citep{Russakovsky2015} task. We had previously found that this model provided the best trade-off between accuracy and complexity~\citep{Jeremie2021}. It also achieves a good model of biological function as measured by the Brain score~\citep{Schrimpf2020}. Compared to other architectures such as ResNet, {\sc Vgg 16} stands out as an ideal candidate. Two notable advantages of the transfer learning method are the robustness and the convergence speed for learning the network. This results in lower total execution time and energy consumption. In~particular, this method allowed us to save computational time during the learning process and thus experiment with several possible strategies (see Figure~\ref{fig:transfer_learning}). We first validated this hypothesis by training a network with random weights: {\sc Vgg SLS} (Supervised Learning from Scratch) as a~control.

During transfer learning, we kept all layers of the {\sc Vgg 16} network, since it is already capable of performing feature extraction on natural scenes, and~re-trained only the last fully connected layer. In~particular, we replaced this last layer trained on {\sc ImageNet} (i.e., with~a vector of dimension $K=1000$ that captures the predicted probability of detection for each of the labels in the {\sc ImageNet} database) with a layer whose output dimension is simply $K=1$ and which represents the predicted probability of detection of a new object of interest, i.e.,~a target rather than a distractor. We then re-trained this fully connected layer, while freezing the weights of the other layers, to~match the pre-trained features (the output of the convolutional layers) with the synset corresponding to the new task implemented by the dataset maker. Following this process, we re-trained a network that we call {\sc Vgg TLC} (Transfer Learning on Classification layers). As~a control, we also tested the effectiveness of freezing the remaining layers by completely re-training all layers of the pre-trained network, {\sc Vgg TLA} (Transfer Learning on All layers). Note that these two networks were trained without any form of data~augmentation.

Since the network is asked to make a binary decision during training (``Is this synset present in this scene?''), we implemented the loss using the binary cross-entropy loss with logits from the {\sc PyTorch} library. We used the stochastic gradient descent (SGD) optimizer from the {\sc PyTorch} library and validated parameters such as batch size, learning rate, and~momentum by performing a sweep of these parameters for each network. During~the sweep, we varied one of these parameters over a given range while leaving the others at their default values  for $25$ epochs. We chose the parameters' values that gave the best average accuracy on the validation set: batch size = $8$, learning rate = $0.00005$, momentum = $0.99$. 
Then, to~increase the generality of our results, we implemented various preprocessing steps on the inputs to introduce more variation into the training dataset: data augmentation. From~the {\sc Vgg TLC} protocol, we tested the effectiveness of this data augmentation using two strategies: first, by~re-training a pre-trained network with a set of custom transformations from the {\sc PyTorch} library: random horizontal flipping (with \emph{p} = 0.5), random vertical flipping (with \emph{p} = 0.5), a~random rotation (\emph{p} = 1), and~random grayscale (with \emph{p} = 0.5), such that we trained the {\sc Vgg TLDA} (Transfer Learning with Data Augmentation) model. Then, the~input images were distorted using the auto-augment function from the {\sc PyTorch} library. This function implements a total of 16 randomly parameterized affine transformations on the inputs to perform data augmentation~\citep{Cubuk2019}, thus defining the {\sc Vgg TLAA} (Transfer Learning with Auto Augment function) model. 
Finally, we studied a {\sc Vgg Random} model trained on the `random' dataset (that is, consisting of two categories defined by randomly chosen labels among the $1000$ labels of {\sc ImageNet}). Note that, although~we implemented all transfer learning strategies on this dataset, as the results were similar for all strategies, we chose to display the networks obtained using the same training protocol as {\sc Vgg TLAA}.


\subsection{Pruning}\label{prunning_methods}
Another network manipulation that we tested is the modification of the CNN architecture. In~particular, we tested the effect of pruning the convolutional layers of the pre-trained network {\sc Vgg 16} to determine the complexity of the features required to categorize a given synset of interest. In~fact, the~{\sc Vgg 16} network can be described as a hierarchically organized pipeline: first, a set of convolutional layers, then a set of fully connected layers~\citep{Simonyan2015}. The~set of convolutional layers is organized into $5$ blocks of $13$ convolutional layers. Within~a block, there is a sequence of convolutional layers followed by a nonlinearity and optionally a normalization (in our case, we did not use or test the batch normalization option). Within~each block, the~image size and the number of channels are constant. In~general, the~resolution decreases from block to block using max-pooling operations, while the number of channels increases from  $64$ at the input to $512$ at the fully connected blocks.

Since the final process of the set of convolutional layers is an adaptive pooling function that produces a characteristic image of constant size equal to $7\times 7$, the~size and architecture of the fully connected layers were kept constant. Therefore, we defined new networks whose names correspond to the number of layers to be pruned. The~network named {\sc Vgg-1} had only its last convolutional layer block pruned, and~then we applied the same learning process as for the network {\sc Vgg TLAA}. We then did the same for the $12$ different depth factors. We have chosen the names of the meshes according to the number of layers removed. Thus, the~network with one layer removed is called ``vgg minus one'', i.e.,~``{\sc Vgg-1}'' (from the deepest {\sc Vgg-1} to the shallowest {\sc Vgg-12}).

\subsection{Accuracy}
Tha accuracy metric will be used to describe the performance of the model. In~effect, the~network is expected to output a binary decision (`Is there an animal in the scene?') and is designed to provide the predicted probability of the presence of a synset of interest in the scene. We considered the output a `target' if the network output was greater than $0.5$ (i.e. $50\%$), otherwise it was considered a `distractor'. A positive true was defined as the case in which the network categorized a `target' if it was a 'target', otherwise it defined a positive false. Similarly, a~true negative was defined when the network categorized a `distractor' when it was indeed a `distractor', otherwise it defined a false negative. Based on these observations, we could determine each time that the networks performed a good categorization and calculate its accuracy as the ratio of the sum of true positives and negatives over the total number of samples tested. To~provide a comparison with the state of the art, we tested the {\sc Vgg 16} and computed its prediction by summing the predictions of the labels belonging to the hyperonymous path of the synsets of interest after the softmax layer, hence {\sc Vgg LUT} (Look Up Table). Accuracy was then computed using the same methodology as for the re-trained networks. We evaluated the accuracy of our different networks on the test set 
using Equation~(\ref{eq:equation1}): 
\begin{equation}
\label{eq:equation1}
\mathrm{Accuracy} = 
\frac{\mathrm{True}_\mathrm{positive}+\mathrm{True}_\mathrm{negative}}
{\mathrm{True}_\mathrm{positive} + \mathrm{True}_\mathrm{negative} + \mathrm{False}_\mathrm{positive} + \mathrm{False}_\mathrm{negative}}
\end{equation}

\vspace{.3cm}
%
\section{Results}
\unskip
\subsection{Performances on Natural Scenes Containing Animals without Transfer~Learning}
\begin{figure}[H]
  \includegraphics[width=1\linewidth]{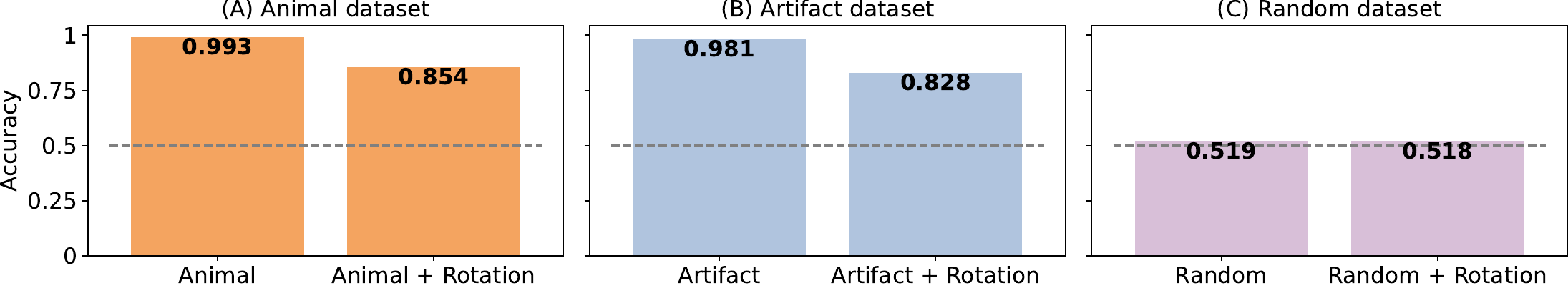}
  \caption{
  \noteLP{One could show some example images which are mis-classified after the rotation.}
  Bar graph representing the accuracy of the {\sc Vgg LUT} network on datasets built with the {(\textbf{A})} `animal', {(\textbf{B})} `artifact', or {(\textbf{C})} control 'random' datasets~\citep{Jeremie2022b}. 
  For~each dataset, the~network is tested with original images (left) or after applying a random rotation (right). The~dotted line represents the chance level for all graphs.}
  \label{fig:figure_pre-trained}
  \end{figure}

Obviously, testing the initial pre-trained net should be one of the first experiments before re-training the neural networks. If~we were to test it on the dataset on which it was trained to categorize an animal, it would indeed perform very well, with a mean accuracy of $0.99$ for categorizing an animal (and $0.98$ for an artifact; see Figure~\ref{fig:figure_pre-trained}). The~goodness of these results is quite stunning compared to human behavioral results and highlights one difference between human and machine~intelligence.

However, as~soon as we added a perturbation such as a random rotation (images similar to~\citep{Guyonneau2006}) into the same dataset, the~performance dropped to $0.85$ for the presence of an animal and $0.83$ for the detection of an artifact, on~par with human performance. Note that if the labels chosen in the task definition have no semantic link, as~is the case for the test on the `random' dataset, the~network cannot perform a correct categorization, with~or without rotation, and~it yields an accuracy close to chance level. 


%
\subsection{Performances on Natural Scenes Containing Animals with Transfer~Learning}
  
We then tested different variations of transfer learning on the task, ``Is there an animal in this visual scene?'', and show the mean accuracies for different datasets, as~summarized in Table~\ref{tab:accuracy}. First, we have seen that the {\sc Vgg LUT} network seems to be robust, as~validated on the dataset used by~\citet{Serre2007}, on~which the network achieves a mean accuracy of $0.95$. Note that it achieves better performances compared to about $0.84$ obtained by the model designed in~\citet{Serre2007} and about $0.80$ in psychophysics, and~this is without any retraining process. Now, let us focus on the network after the transfer learning process, as~the {\sc Vgg TLC}, {\sc Vgg TLA}, {\sc Vgg TLDA}, and~{\sc Vgg TLAA} reached similar levels of performance on the test set (with $0.97$, $0.96$, $0.97$, and $0.95$, respectively) and also maintained robust categorization on the~\citet{Serre2007} dataset (with $0.94$, $0.92$, $0.91$, and $0.88$, respectively). Compared to the {\sc Vgg SLS}, which could only reach $0.64$ on the same task, these results show that transfer learning allows us to obtain highly accurate networks for the categorization of a synset of interest. Note that this low performance is only due to the computational limits that we imposed in our study. We then focused on the robustness of the categorization of the different data augmentation strategies ({\sc Vgg TLC}, {\sc Vgg TLA}, {\sc Vgg TLDA}, and {\sc Vgg TLAA}) compared to the state of the art {\sc Vgg LUT} and the expected performance in neurobiological~models. 
\begin{table}[H]
\setlength{\tabcolsep}{2.9mm}
\caption{{Mean} 
 accuracies for the ultrafast image categorization of an animal in a scene from a newly built dataset using our library dataset maker with the synset `animal' (top) or for the~\citet{Serre2007} dataset (bottom) for different transfer learning~strategies. }
\label{tab:accuracy}
\begin{tabularx}{\textwidth}{c c c c c c c c}
 \toprule
 \multicolumn{8}{c}{\textbf{Dataset Built with Dataset Maker, Synset: Animal}} \\
 \midrule
 & LUT  & TLC & TLA & TLDA & TLAA & SLS &  \\ [0.5ex] 
 \midrule
 Accuracy & 0.99  & 0.97 & 0.96 & 0.97 & 0.95 & 0.64 &  \\ 
 \midrule
 \multicolumn{8}{c}{\textbf{Dataset from~\citet{Serre2007}}} \\
 \midrule
 & LUT  & TLC & TLA  & TLDA & TLAA & Serre et al.'s model & Human\\ [0.5ex] 
 \midrule
 Accuracy & 0.95 & 0.94 & 0.92  & 0.91 & 0.88 & 0.84 & 0.80\\ 
 \bottomrule
 \end{tabularx}
\end{table}

%
\subsection{Robustness of the Categorization with Different Geometric~Transformations}

\begin{figure}[H]
\includegraphics[width=1\linewidth]{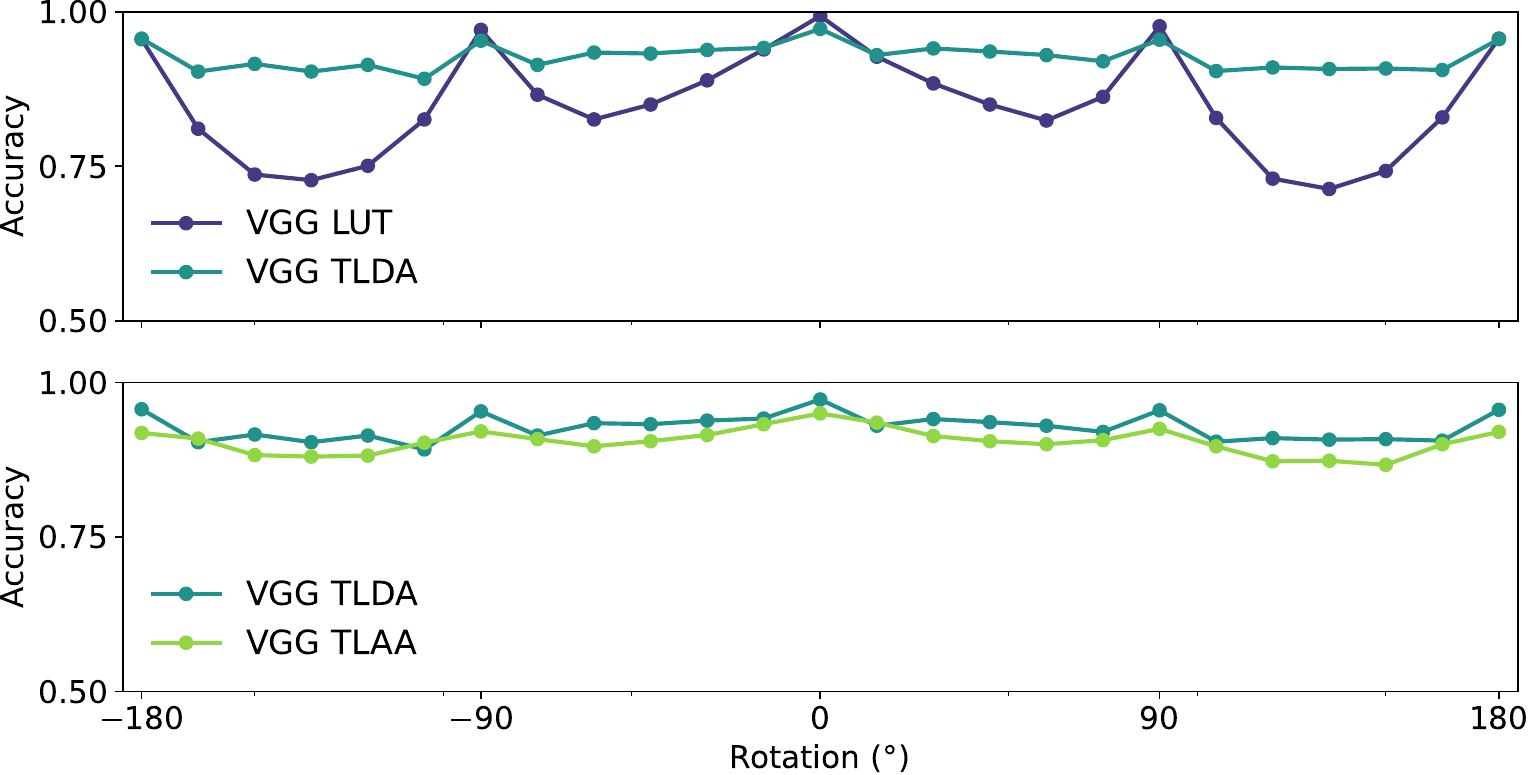}
\caption{%
Average accuracy in the test dataset of the different networks for different rotations of the input image. (Top) The {\sc Vgg LUT} displayed with the {\sc Vgg TLDA} networks. (Bottom) The {\sc Vgg TLAA} displayed with the {\sc Vgg TLDA} networks. The~rotation is applied around the center with an angle ranging from $-180^{\circ}$ to $+180^{\circ}$.
\label{fig:rotation}}
\end{figure}

Since we were looking for the best robustness for this task, we tested {\sc Vgg TLC}, {\sc Vgg TLA}, {\sc Vgg TLDA}, {\sc Vgg TLAA}, and~{\sc Vgg LUT} on the newly constructed dataset using our dataset maker library with the synset `animal'. We applied either a grayscale filter or a~vertical or a horizontal reflection to the input (see Table~\ref{tab:accuracy2}). We also tested the robustness to rotation by rotating the image around the center by an angle ranging from $-180^{\circ}$ to $+180^{\circ}$ (see Figure~\ref{fig:rotation}). All these networks maintained good average accuracy on the returned dataset and on the grayscale dataset (see Table~\ref{tab:accuracy2}). These results were consistent with psychophysical results showing that ultrafast categorization is robust to a grayscale transformation~\citep{Zhu2013}. Only {\sc Vgg TLDA} and {\sc Vgg TLAA} seemed to show robust accuracy at all angles, with~peaks in accuracy at the cardinal orientations ($-180^{\circ}$, $-90^{\circ}$, $0^{\circ}$, $90^{\circ}$, and~$180^{\circ}$), which could be explained by the pre-training weights of the networks, as~they correspond to the peaks found in the categorization of {\sc VGG 16}. We conclude here that data augmentation provides a more robust categorization of the synset of interest by the network, as~the {\sc Vgg TLDA} and {\sc Vgg TLAA} achieve better performance in this task. In~addition, the~protocol used to re-train the network {\sc Vgg TLAA}, with~the auto-augment function of the library {\sc PyTorch}~\citep{Cubuk2019}, is also better than our custom data augmentation. 
The~performance of {\sc Vgg TLAA} is very close to that of {\sc Vgg TLDA}, with a tendency for {\sc Vgg TLAA} to be more robust to rotation. Therefore, the~{\sc Vgg TLAA} network is the best fit for psychophysical observations due to its stability and robustness of categorization to different image transformations~\citep{Rousselet2003, Guyonneau2006}. In the following, we therefore focused on exploring the features that this model relies on to perform its categorization.

\begin{table}[H]
\setlength{\tabcolsep}{6.9mm}
\caption{{Mean} 
 accuracies for ultrafast image categorization of an animal in a scene  using various geometric transformation on the input: vertical flip, horizontal flip, grayscale filter. These transformations were implemented using our dataset maker library with the synset `animal' for four re-trained networks: {\sc Vgg TLC}, {\sc Vgg TLA}, {\sc Vgg TLDA}, and {\sc Vgg TLAA}. It was compared with the state-of-the-art network {\sc Vgg LUT}. All the transformations used here were performed using the {\sc PyTorch} library~\citep{Paszke2019}.}
\label{tab:accuracy2}
\begin{tabularx}{\textwidth}{c c c c c c}
 \toprule
  & \textbf{LUT}  & \textbf{TLC} & \textbf{TLA} & \textbf{TLDA} & \textbf{TLAA}\\ 
 \midrule
 \multicolumn{6}{c}{Vertical flip} \\
 \midrule
 Accuracy & 0.96 & 0.94 & 0.94 & 0.95 & 0.93\\ 
 \midrule
 \multicolumn{6}{c}{Horizontal flip} \\
 \midrule
 Accuracy & 0.99 & 0.97 & 0.96 & 0.97 & 0.95\\ 
 \midrule
 \multicolumn{6}{c}{Grayscale filter} \\
 \midrule
 Accuracy & 0.96 & 0.95 & 0.93 & 0.95 & 0.93\\ 
 \bottomrule
 \end{tabularx}
\end{table}
\unskip
%

%

\subsection{What Features Are Necessary to Achieve the Task?}

\begin{figure}[H]
\includegraphics[width=\linewidth]{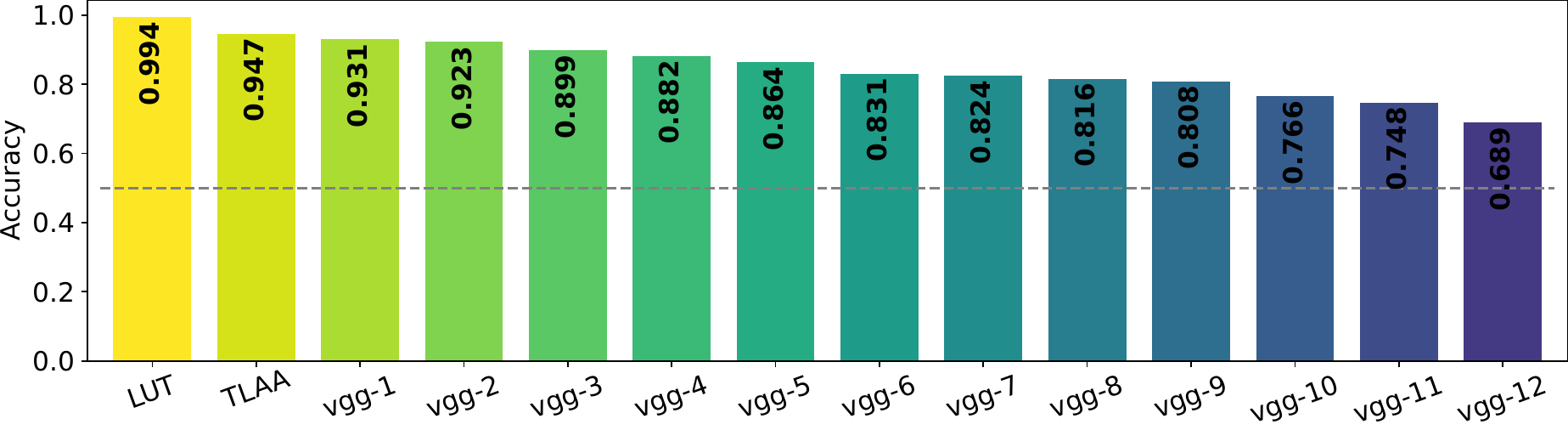}
\caption{
Average accuracy obtained as a function of depth in pruning networks. The~networks are re-trained to categorize animals and tested on test datasets based on {\sc ImageNet}  images constructed using the `animal' synset. The prefix ``vgg-'' indicates the number of convolutional layer blocks pruned from the original network (see Section~\ref{prunning_methods} for more details). The~dotted line represents the chance level for all graphs.}
\label{fig:pruning}
\end{figure}

We designed an experiment in which we gradually removed layers from a pre-trained network {\sc Vgg 16} for $12$ different ``depth'' factors. For~each level, we tested the re-trained pruned networks to categorize an animal in a scene for our dataset {\sc ImageNet}. {\sc Vgg LUT} and {\sc Vgg TLAA} achieved the best accuracy for this task (see Figure~\ref{fig:pruning}). The~accuracies of the networks remained similar to the performance found by~\citet{Serre2007} with~a slight drop between {\sc Vgg-9} and {\sc Vgg-12}. This is not a surprise, as their model relied on low-level features~\citep{Perrinet2015}. Note that the computational time required to perform the categorization decreased with the depth of the network (in seconds on a Quadro RTX 5000 GPU, we obtained {\sc Vgg TLAA} $= 0.005 \pm 0.0001$ and {\sc Vgg-8} $= 0.003 \pm 0.0001)$ (see Table~\ref{tab:mean_time}).
\noteLP{Performances based on low-level features: one could easily add an experiment where one adds a transformation that keeps only the contours of the objects - a paper = https://www.nature.com/articles/s41598-019-43956-3}
%

\begin{table}[H]
\caption{{Table} 
 showing the average time in seconds required for networks to perform a prediction for a $256 \times 256$ resolution image with a Quadro RTX 5000~GPU. }
\label{tab:mean_time}
\begin{tabularx}{\textwidth}{c c c c c}
 \toprule
 \multicolumn{5}{c}{\textbf{Mean Time (s)}} \\
 \midrule
{\sc Vgg TLAA} & {\sc Vgg-1} & {\sc Vgg-2} & {\sc Vgg-3} & {\sc Vgg-4}\\ 
 \midrule
$0.0057 \pm 0.0001$ & $0.0055 \pm 0.0005$ & $0.0054 \pm 0.0008$ & $0.0051 \pm 0.0003$ & $0.0047 \pm 0.0002$\\ 
 \midrule
 {\sc Vgg-5} & {\sc Vgg-6} & {\sc Vgg-7} & {\sc Vgg-8} & {\sc Vgg-9}\\ 
 \midrule
$0.0043 \pm 0.0001$ & $0.0035 \pm 0.0007$ & $0.0031 \pm 0.0001$ & $0.0027 \pm 0.0003$ & $0.0022 \pm 0.0002$ \\
  \midrule
{\sc Vgg-10} & {\sc Vgg-11} & {\sc Vgg-12} & & \\ 
 \midrule
$0.0018 \pm 0.0007$ & $0.0013 \pm 0.0006$ & $0.0008 \pm 0.0006$ &  & \\ 
 \bottomrule
 \end{tabularx}
\end{table}

\begin{figure}[H]
\includegraphics[width=1\linewidth]{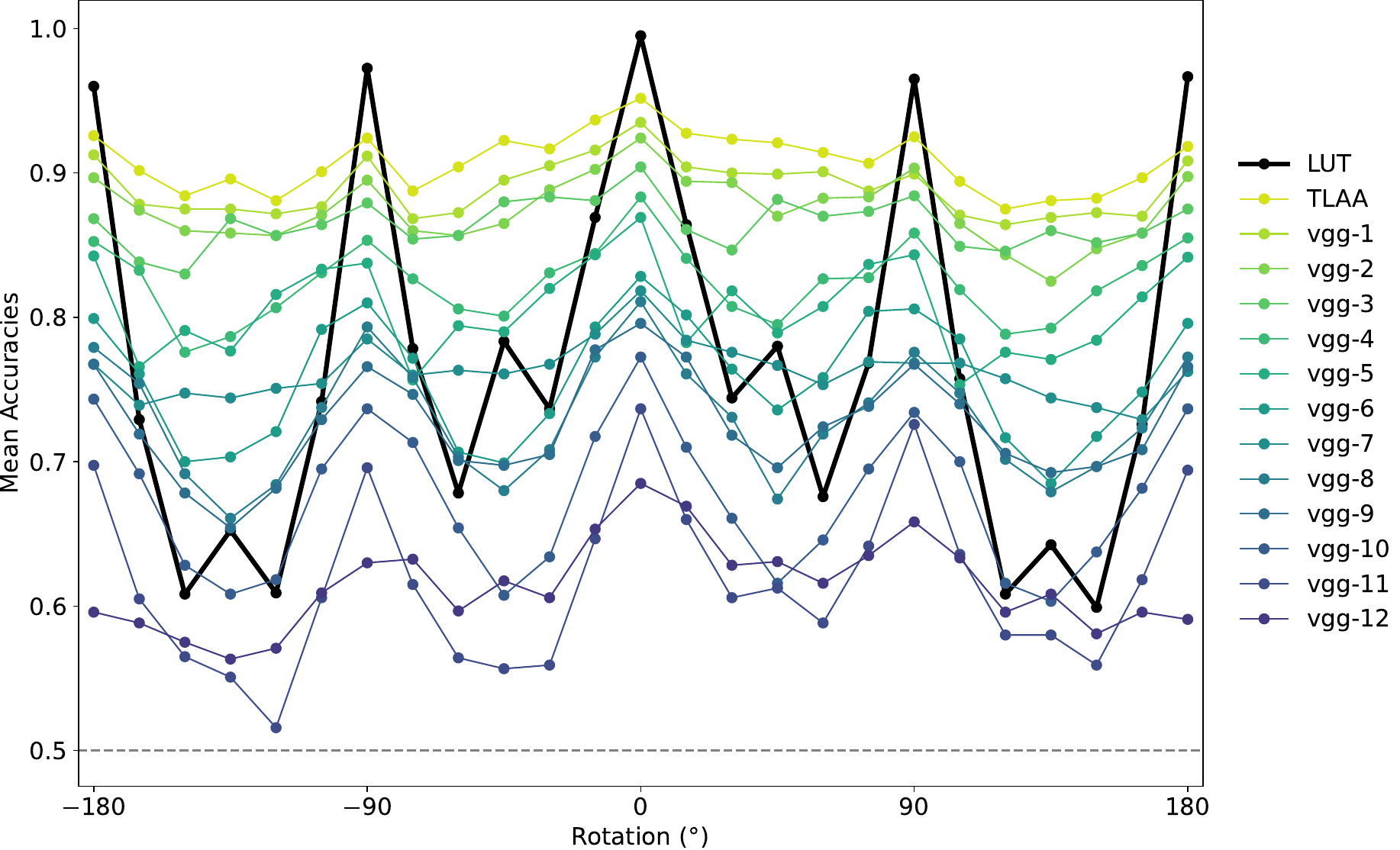}
\caption{
Average accuracy over the test dataset of the re-trained networks for rotations around the center from $-180^{\circ}$ to $+180^{\circ}$. These networks were tested on a dataset based on {\sc ImageNet} images constructed using the `animal' synset. The~index after ``vgg-'' indicates the number of convolutional layers pruned in the networks (with vgg-1 being the deepest and vgg-12 being the shallowest).
}
\label{fig:pruning_robustness}
\end{figure}

We also tested all pruned networks on our {\sc ImageNet} dataset by rotating the image around the center from $-180^{\circ}$ to $+180^{\circ}$; however, the~categorization may lose robustness with fewer layers (see Figure~\ref{fig:pruning_robustness}). Indeed, as~the number of layers and the mean accuracy after rotation decreased, the~standard deviation of the mean accuracy increased ({\sc Vgg TLAA} $= 0.91 \pm 0.02$, {\sc Vgg-8} $= 0.73 \pm 0.04$). Although~the networks seem to be able to categorize an animal with fewer layers, they seem to trade this advantage for a lower robustness to transformations such as~rotations.

\begin{figure}[H]
\includegraphics[width=1\linewidth]{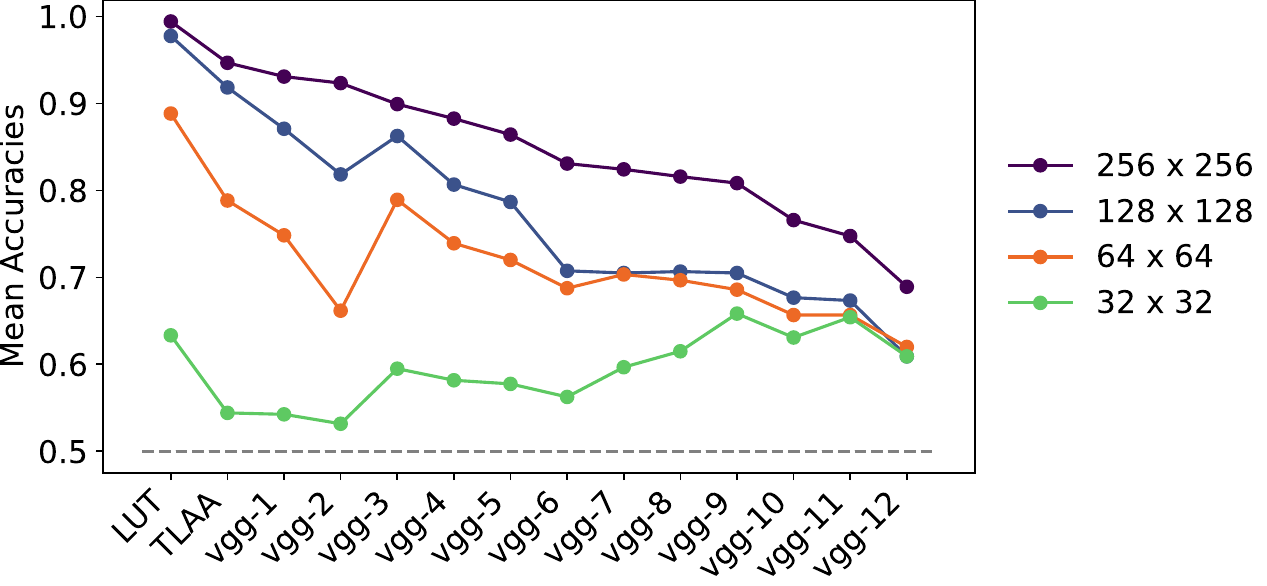}
\caption{
Average accuracy obtained for the differentially pruned networks over the `shuffled' test dataset, where we applied a shuffled transformation to the input image. We show the results as we decreased the size of the shuffled patches on the images. The~networks were retrained to categorize animals and tested on datasets based on {\sc ImageNet} images created using the `animal' synset. The~index after ``vgg-'' indicates the number of convolutional layers pruned in the networks. The~dotted line represents the chance level for all plots.
}
\label{fig:shuffle} 
\end{figure}

To get a better idea of the size of the feature maps needed to categorize an animal in a scene, we tested the networks on a new ``shuffled'' dataset, where the image had been divided into square patches of different sizes and then blended to generate a new image~\citep{Biederman1972}. Since CNN networks are by definition robust to translation, patch translation should have minimal impact on categorization unless it breaks some necessary patterns in the images. With~few layers, the~networks should rely on low-level features to perform their categorization, and~indeed we obtained an idea of the size of feature maps required for different depths. In~fact, between~patch sizes $256\times 256$ and $64\times 64$, the~categorization of the networks was robust to this transformation (see Figure~\ref{fig:shuffle}). However,~as soon as we reached the patch size of $32\times 32$ pixels, the~accuracy of all networks dropped sharply. Furthermore, there seemed to be a transition between deeper and medium networks, as~the latter gave better average accuracies for this task. As~a consequence, the~size of the feature maps needed to perform such a task varies with the depth of the network. For~example, {\sc Vgg TLAA} appears to rely on feature map sizes between $32\times 32$ and $64\times 64$ pixels, as~its accuracy drops when we exceed this threshold (see Figure~\ref{fig:shuffle}); however, further study is needed to quantify this feature map size. In~a future application, we could extract feature maps from these low-level layers to better understand the features needed to perform this task. This would allow us to design a stimulus set for a psychological task such as in~\citet{Thorpe1996}. Such a test could be relevant to whether these features are sufficient to categorize an animal in a flashed~scene.

\unskip

\subsection{Dependence of Accuracy Scores between the Two~Tasks}

\begin{figure}[H]
  \includegraphics[width=1\linewidth]{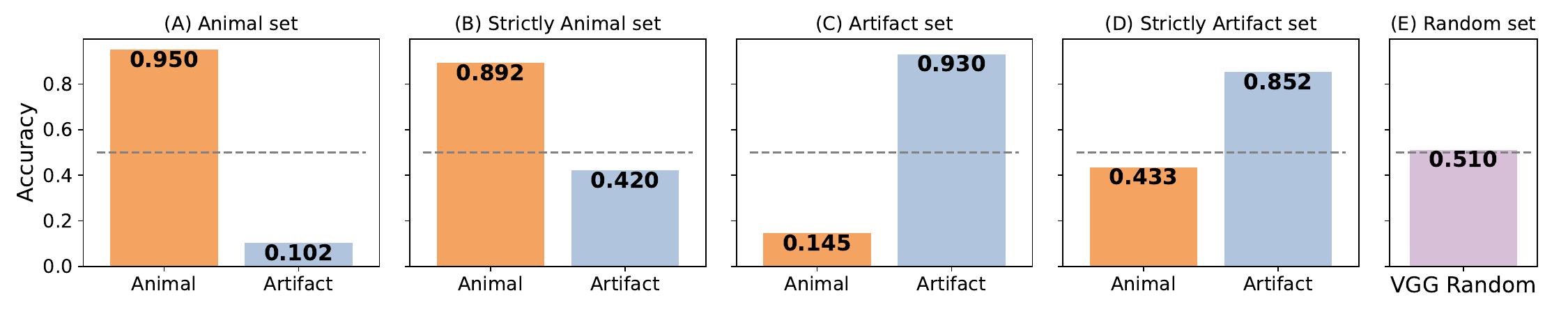}
  \caption{Bar graph representing the accuracy of networks re-trained to categorize an animal: {\sc Vgg Animal} (in orange); re-trained to categorize an artifact: {\sc Vgg Artifact} (in blue); or re-trained to categorize a random distribution of synset: {\sc Vgg Random} (in violet); (\textbf{A}) on the dataset built with the `animal' synset; (\textbf{B}) on the dataset built with the `animal' synset where the distractor ``is not an artifact'' thus constitutes the ``strictly animal'' dataset; (\textbf{C}) on the dataset built with the `artifact' synset; (\textbf{D}) on the dataset built with the `artifact' synset where the distractor ``is not an animal'' thus constitutes  the ``strictly artifact'' dataset; (\textbf{E}) on a dataset built with a random distribution of synset (see Figure~\ref{fig:transfer_learning}). The~dotted line represents the chance level for all~graphs. 
}
  \label{fig:animal_artifact}
  \end{figure}

We examined dependence of learning performance of {\sc Vgg TLAA} between two tasks by introducing a variation of the synset of interest in the construction of the dataset. We used our dataset maker tool with the keyword `artifact', thus generating a new network trained to categorize the presence of the `artifact' synset in a natural scene: {\sc Vgg Artifact}. We displayed the average accuracy of the networks trained to detect the `animal' synset (here {\sc Vgg Animal} stand for our {\sc Vgg TLAA}) on the dataset constructed with the `animal' synset (respectively trained to detect the artifact synset tested on the dataset constructed with the artifact synset). Next, we tested the networks trained to detect the `animal' synset on the dataset constructed with the `artifact' synset and vice~versa. Here, by~exposing the predictions for the `animal' and `artifact' synsets, we highlight a bias in the composition of the dataset. Although~the outputs are independent, the~`animal' images confidently match the `non-artifact' images (and vice~versa), thus facilitating global detection (see Figure~\ref{fig:animal_artifact}A,B).

To infer the influence of this bias on the performance of the network, we generated through the dataset maker a dataset based on the `animal' synset where, in~addition to not being animals, the~distractors would also not be `artifacts'. This defines the `strictly animal' set (respectively, one  defines the `strictly artifact' set based on the `artifact' synset where, in~addition to not being an artifact, the~distractors would also not be an `animal'). Once this distinction was made, although~there is a loss in performance for both networks, they remained fit for their respective tasks by maintaining an accuracy above $0.8$. On~the other hand, they did not seem to be able to predict the absence of their respective sentences once the ensemble was modified (see Figure~\ref{fig:animal_artifact}B,D). These results reinforce the argument that, despite task independence, the~composition of the dataset can generate bias in network~categorization. 

\begin{figure}[H]
    \subfloat[\centering{Synset Animal \\ Target: Human fails}]{\includegraphics[width=.25\linewidth]{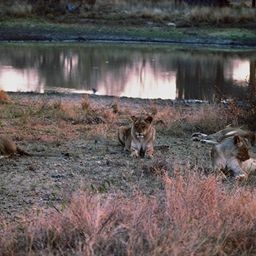}}\hspace{15pt}
    \subfloat[\centering{Synset Animal \\ Target: Both fail}]{\includegraphics[width=.25\linewidth]{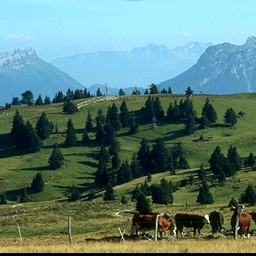}}\hspace{15pt}
    \subfloat[\centering{Synset Animal \\ Target: TLAA fails}]{\includegraphics[width=.25\linewidth]{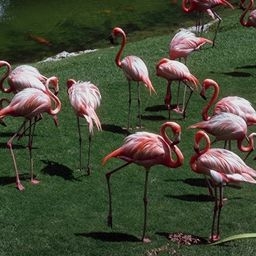}}%
    \\
    \subfloat[\centering{Synset Animal \\ Distractor: Human fails}]{\includegraphics[width=.25\linewidth]{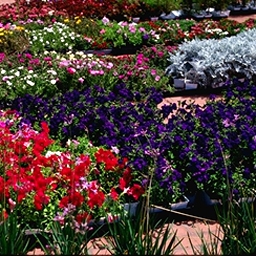}}\hspace{15pt}
    \subfloat[\centering{Synset Animal \\ Distractor: Both fail}]{\includegraphics[width=.25\linewidth]{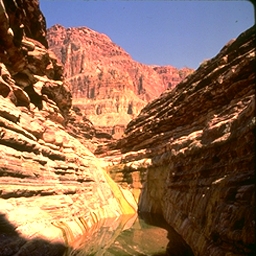}}\hspace{15pt}
    \subfloat[\centering{Synset Animal \\ Distractor: TLAA fails}]{\includegraphics[width=.25\linewidth]{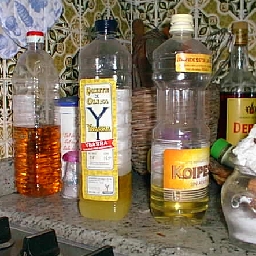}}%
    \caption{
    Some prototypical example images where either (\textbf{A})~humans failed to categorize an animal (or in (\textbf{D}), the absence of an animal) in a scene but our model succeeds; (\textbf{B})~humans and our model both failed to categorize an animal ((\textbf{E}), the absence of an animal) in the scene; (\textbf{C})~the model failed to categorize the presence of an animal ((\textbf{F}), the absence of an animal) in a scene but the human succeeded. The~psychophysical data and the images were taken from the dataset used by~\citet{Serre2007}.
    \label{fig:animal_artifact_examples}}
    \end{figure}

As a control, we tested the {\sc Vgg Random} network on the corresponding dataset (see Section~\ref{dataset_maker} for details). As~it obtains an average accuracy close to the one obtained with the {\sc Vgg SLS} network, its poor performance can be explained by the fact that the pre-trained weights of the {\sc Vgg 16} network do not match the new task. 
Incidentally, this bias is also present in the dataset used by~\citet{Serre2007}. However, when we compared the performance of the humans on this dataset with the performance achieved by the network on a frame-by-frame basis, we found a high correspondence (about $0.84$) in their correct predictions. Indeed, for~some images, the~networks failed at categorizing but~the human succeeded, and vice~versa. For~some images, both the network and the human succeeded or failed in categorizing an animal, and there were cases where the network was wrong but the humans responded correctly on average (see Figure~\ref{fig:animal_artifact_examples}). We have displayed images where one human or both a human and our model failed to categorize an animal in the scene, as~this may reflect the specific features that humans or our models rely on to perform their categorization. This close relationship between human and network responses could allow us to select images and design physiological and psychophysical tests to infer the features necessary for such~detection.

%
\section{Discussion}

In this paper, we have shown that we can re-train networks using transfer learning to apply them to an ecological image categorization task and obtain insights on visuo-cognitive processes. Such outcomes could in particular be beneficial when studying impaired systems such as in Autism Spectrum Disorder~\citep{Vanmarcke2016}. These artificial networks achieve accuracies similar to those found in psychophysical responses in humans. In~the image processing flow at work in convolution networks, the~position of the feature maps has no influence on the activation of receptive fields. Since translation is a shift in the position of the feature maps, these networks are supposed to be robust to translation. However,~a transformation by a rotation constitutes then a global perturbation of the features composing the maps. Thus, since the features are different, rotation can lead to the solicitation of different receptive fields. If~these new receptive fields are not previously learned, the~network will be unable to generalize. This could explain the differences in performance between learning protocols involving or not involving rotations. Furthermore, the~robustness of the categorization is comparable to that found in psychophysical data. In~particular, we have shown quantitatively that the categorization of the re-trained networks may be robust to transformations such as rotations, reflections, or grayscale filtering, such as is observed in humans~\citep{Thorpe1996,Rousselet2003}.

We have studied networks that learn to detect if an image contains an animal or an artifact. Two independent networks each re-trained on each of the two categorization tasks used to highlight a link or~rather a bias in this categorization. This kind of bias is also found in humans and seems to impact the categorization as well~\citep{Bogadhi2020} and could be linked to top--down influences~\citep{Xu2019}. The~question of detecting an animal in an image is indeed tightly linked to that of detecting an artifact, allowing for the possibility of the less likely appearances of an animal object (like a teddy bear) or of a non-animal non-object (like a mountain). The~study of this kind of bias could possibly allow for building ecologically-relevant datasets to maximize the learning process of the networks in order to discover more about the features needed for categorization~\citep{Mehrer2021}.

While the level of $80\%$ correct categorization between humans and machines in this type of task is similar, both could be driven to make different ``mistakes'', and~these particular examples could then be used as subjects for studies in the design of psychophysical tests. In~addition, these systematic errors could be a window into some processes in our understanding of primate visual pathways. The~last part of our study was based on the search for the features necessary for categorization. We found that, in~agreement with the studies of~\citet{Serre2007}, a~simple feed-forward network based on low-level features was sufficient to perform categorization efficiently. Moreover, we estimated the size of the features needed to be about $32\times 32$ pixels and $64\times 64$ pixels. Although~categorization is still possible at this very low computational cost, we quantitatively show that it gradually loses robustness. 
%


\noteLP{parler de Grandmother Cells and Distributed Representations~\citep{Thorpe2011} }


\section{Perspectives}

\begin{figure}[H]
\includegraphics[width=1\linewidth]{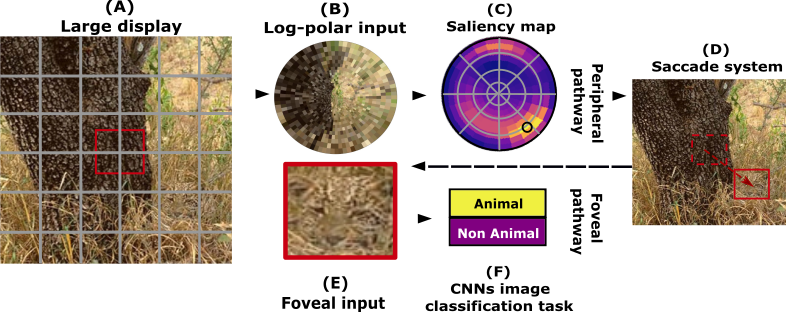}
\caption{
Model built over the anatomical visual processing pathways observed in mammals, namely the ``what'' and the ``where'' pathways: the peripheral pathway (top row) is applied to a large display from a natural scene (\textbf{A}): it is first transformed into a retinotopic log-polar input (\textbf{B}), and we then learn to return a ``saliency map'' (\textbf{C}). The~latter infers, for~different positions in the target, the~predicted accuracy value that can be reached by the foveal pathway, mimicking the ``where'' pathway used for global localization. The~position with the best accuracy will feed a saccade system (\textbf{D}), adjusting the fixation point at the input of the foveal pathway (bottom row). It takes a subsample (\textbf{E}) of the large display (\textbf{A}), over~which a categorization is done (\textbf{F}), mimicking the ``what'' pathway.
}
\label{fig:fast_and_curious}
\end{figure}

One of the main goals of this study was to provide a comparison for an ecological and well-studied task used in visual neuroscience. Although~this study focuses on the analysis of categorization, it is a necessary step for a well-known task in the field of vision: visual search. This task consists of the simultaneous localization and detection of a visual target of interest. Applied to the case of natural scenes, visually searching, for example, for an animal (either prey, a~predator, or a partner) constitutes a challenging problem due to large variability over the  numerous visual dimensions. Previous models managed to solve the visual search task by dividing the image into sub-areas. This is at the cost, however, of~computer-intensive parallel processing on relatively low-resolution image samples~\citep{Liu2016,Ren2016}. Taking inspiration from natural vision systems~\citep{Mishkin1983}, we developed a model that was built over the anatomical visual processing pathways observed in mammals, namely the ``what'' and the ``where'' pathways~\citep{Dauce2020a}. It operates in two steps; one by selecting a region of interest, before~knowing its actual visual content, through an ultrafast/low resolution analysis of the full visual field, and~the second providing a detailed categorization of the detailed ``foveal'' selected region attained with the saccade~\citep{Yarbus1961} (see Figure~\ref{fig:fast_and_curious}). In~this perspective, our work would be a deepening of the knowledge and models necessary for the realization of the ``what'' pathway. Modeling this dual-pathways architecture allows for offering an efficient model of visual search as active vision. In~particular, it allows us to fill the gap with the shortcomings of CNNs with respect to physiological performances~\citep{New2007}. In~the future, we expect to apply this model to better understand visual pathologies in which there exists a deficiency of one of the two pathways~\citep{Wiecek2012} while contributing to the field of computer vision.

\unskip

\vspace{6pt}

\authorcontributions{
Conceptualization, J.-N.J. and L.U.P.; methodology, J.-N.J.; software, J.-N.J.; validation, J.-N.J. and L.U.P.; formal analysis, J.-N.J. and L.U.P.; investigation, J.-N.J. and L.U.P.; resources, J.-N.J. and L.U.P.; data curation, J.-N.J. and L.U.P.; writing---original draft preparation, J.-N.J. and L.U.P.; writing---review and editing, J.-N.J. and L.U.P.; visualization, J.-N.J.; supervision, L.U.P.; project administration,  L.U.P.; funding acquisition, L.U.P. All authors have read and agreed to the published version of the manuscript.
}

\funding{
  Authors received {funding} 
  from the Agence Nationale de la Recherche project number {ANR-20-CE23-0021} 
  (``\href{https://laurentperrinet.github.io/grant/anr-anr/}{AgileNeuroBot}'', accessed on 15 March 2023) and from the french government under the France 2030 investment plan, as part of the Initiative d'Excellence d'Aix-Marseille Universit\'{e} - A*MIDEX grant number {AMX-21-RID-025} 
   ``\href{https://laurentperrinet.github.io/grant/polychronies/}{Polychronies}'', accessed on 15 March 2023. 
}

\institutionalreview{Not applicable.} 

\informedconsent{Not applicable.} 

\dataavailability{\DataAvailability } 

\acknowledgments{\Acknowledgments }

\conflictsofinterest{The authors declare no conflict of~interest.} 



\clearpage
\abbreviations{Abbreviations}{
\noindent 
\begin{tabular}{@{}ll}
(D)CNN & (Deep) Convolutional Neural Network\\
LUT & Look Up Table\\
MNIST & Modified or Mixed National Institute of Standards and Technology\\
SLS & Supervised Learning from Scratch\\
TLA & Transfer Learning on All layers\\
TLAA & Transfer Learning with Auto Augment function\\
TLC & Transfer Learning on Classification layers\\
TLDA & Transfer Learning with Data Augmentation\\
VGG & Vision Geometry Group
\end{tabular}
}

\appendixtitles{no} 

\noteLP{we could add some supplementary methods from the notebook: further  test on generated images with LDM, further controls, ...}


\begin{adjustwidth}{-\extralength}{0cm}

\reftitle{References}






\PublishersNote{}
\end{adjustwidth}
\end{document}